\pdfoutput=1

\documentclass[11pt,twoside,a4paper,cmspaper,final,collab]{cms-tdr}
\def\svnVersion{49cfed8-D}\def\svnDate{2020/05/22}\def\cmsCernNoTag{CERN-EP-2019-226}\def\cmsCernDate{\today}\def\cmsMessage{"Published in Physical Review Letters as \href{http://dx.doi.org/10.1103/PhysRevLett.124.202001}{\doi{10.1103/PhysRevLett.124.202001}.}"}

\begin{document}\cmsNoteHeader{TOP-19-005}

\hyphenation{had-ron-i-za-tion}
\hyphenation{cal-or-i-me-ter}
\hyphenation{de-vices}
\newlength\cmsFigWidth
\ifthenelse{\boolean{cms@external}}{\setlength\cmsFigWidth{0.45\textwidth}}{\setlength\cmsFigWidth{0.49\textwidth}}
\ifthenelse{\boolean{cms@external}}{\providecommand{\cmsLeft}{upper\xspace}}{\providecommand{\cmsLeft}{left\xspace}}
\ifthenelse{\boolean{cms@external}}{\providecommand{\cmsRight}{lower\xspace}}{\providecommand{\cmsRight}{right\xspace}}

\newcommand{\mjet}{\ensuremath{m_\text{jet}}\xspace}
\newcommand{\mtop}{\ensuremath{m_{\cPqt}}\xspace}
\newcommand{\ptlep}{\ensuremath{\pt^\ell}\xspace}
\newcommand{\ptjet}{\ensuremath{\pt^\text{jet}}\xspace}
\newcommand{\antikt}{anti-\kt}
\newcommand{\ptrel}{\ensuremath{p_\text{T, rel}}\xspace}

\newcommand{\muf}{\ensuremath{\mu_{\text{F}}}\xspace}
\newcommand{\mur}{\ensuremath{\mu_{\text{R}}}\xspace}
\newcommand{\ljets}{\ensuremath{\ell\text{+jets}}\xspace}
\newcommand{\Wjets}{\ensuremath{\PW\text{+jets}}\xspace}
\newcommand{\pp}{\ensuremath{\Pp\Pp}\xspace}
\providecommand{\jet}{{\HepParticle{j}{}{}}\Xspace}
\newcommand{\fb}{\unit{fb}}
\newcommand{\model}{\ensuremath{\,\text{(model)}}\xspace}
\newcommand{\exper}{\ensuremath{\,\text{(exp)}}\xspace}

\newlength\cmsTabSkip\setlength{\cmsTabSkip}{1ex}

\cmsNoteHeader{TOP-19-005}
\title{Measurement of the jet mass distribution and top quark mass in hadronic decays of boosted top quarks in  \texorpdfstring{\pp collisions at $\sqrt{s} = 13\TeV$}{pp collisions at sqrt(s)=13 TeV}}

\date{\today}

\abstract{
   A measurement is reported of the jet mass distribution
   in hadronic decays of boosted top quarks
   produced in \pp collisions at $\sqrt{s}=13$\TeV.
   The data were collected with the CMS detector at the LHC and
   correspond to an integrated luminosity of 35.9\fbinv.
   The measurement is performed in the lepton+jets channel
   of \ttbar events, where the lepton is an electron or muon.
   The products of the hadronic top quark decay
   $\cPqt \to \cPqb\PW \to \cPqb \cPq \cPaq'$
   are reconstructed as a single jet
   with transverse momentum larger than 400\GeV.
   The $\ttbar$ cross section as a function of the jet mass is unfolded
   at the particle level and used to extract a value of the top quark
   mass of $172.6 \pm 2.5\GeV$.
   A novel jet reconstruction technique is used for the first time at the LHC,
   which improves the precision by a factor of three relative to an earlier measurement. 
   This highlights the potential of measurements using boosted top quarks, 
   where the new technique will enable future precision measurements.
}

\hypersetup{
pdfauthor={CMS Collaboration},
pdftitle={Measurement of the jet mass distribution and top quark mass in hadronic decays of
boosted top quarks in pp collisions at 13 TeV},
pdfsubject={CMS},
pdfkeywords={CMS, physics, top quark, jet mass}}

\maketitle

The top quark is the most massive known elementary particle. Its large mass \mtop leads to significant
contributions from quantum corrections to the mass of the Higgs boson and precision observables
in the electroweak sector.
As a consequence, the top quark plays an important role in the mechanism of electroweak symmetry breaking.
Precision measurements of \mtop provide a crucial input for consistency checks
of the standard model~\cite{Haller:2018nnx, PDG:2018}.
Direct measurements of \mtop at the CERN LHC
reach a precision of around 0.5\GeV~\cite{Aaboud:2016igd, Aaboud:2017mae, Aaboud:2018zbu,
Khachatryan:2015hba, Sirunyan:2017idq, Sirunyan:2018gqx, Sirunyan:2018mlv}.
However, an ambiguity in the interpretation of the results
originates from the modeling of parton-shower dynamics and nonperturbative effects in
quantum chromodynamics (QCD).
The result can depend on the Monte Carlo (MC) event generator, the tuning of its
free parameters, and on the observables used~\cite{Ravasio:2018lzi}.
Precisely relating the experimentally obtained value of \mtop to the pole mass or a
mass in another well-defined renormalization scheme is
therefore difficult from first principles~\cite{Hoang:2018zrp}.

As an alternative, a value of the pole mass can be extracted through
measurements of the total~\cite{Aad:2014kva, Khachatryan:2016mqs, Sirunyan:2018goh, Sirunyan:2018goh, Aad:2019hzw}
and differential~\cite{Aaboud:2017ujq, Sirunyan:2019zvx} \ttbar production cross sections,
with a precision of approximately 1\GeV.
These measurements are dominated by \ttbar threshold production, where uncertainties due to
parton distribution functions (PDFs) and higher-order QCD corrections
are important~\cite{Kidonakis:2014isa, Guzzi:2014wia, Muselli:2015kba}.
Another way to determine \mtop involves measuring top quarks produced with
large Lorentz boosts, where the decay products $\cPqt \to \cPqb \PW \to \cPqb \qqbar'$
are contained in a single jet.
The jet mass (\mjet) peak location is sensitive to
\mtop, and can be calculated from first
principles~\cite{Fleming:2007qr, Fleming:2007xt, Jain:2008gb, Hoang:2015vua,
Butenschoen:2016lpz, Hoang:2017kmk, Hoang:2019fze}
in soft-collinear effective theory~\cite{Bauer:2000ew, Bauer:2000yr, Bauer:2001ct, Bauer:2001yt}.

A past measurement reporting the $\ttbar$ cross section as a function of \mjet in the $\ell$+jets final state,
where $\ell$ is an electron or muon,
was carried out in proton-proton (\pp) collisions at
$\sqrt{s}=8\TeV$~\cite{Sirunyan:2017yar}.
This Letter reports a new measurement of the \mjet distribution
in \pp collisions at 13\TeV using several important improvements,
including jet clustering with the XCone algorithm~\cite{Stewart:2015waa},
used for the first time in an LHC analysis, and an improved
unfolding procedure using sideband regions with high granularity.

The central feature of the CMS apparatus is a superconducting solenoid of 6\unit{m} internal diameter,
providing a magnetic field of 3.8\unit{T}. A silicon pixel and strip tracker, a lead tungstate crystal
electromagnetic calorimeter (ECAL), and a brass and scintillator hadron calorimeter,
each composed of a central barrel and two end sections, reside within the solenoid volume.
Forward calorimeters extend the pseudorapidity ($\eta$) coverage provided by the barrel and end detectors.
Muons are detected in gas-ionization chambers embedded in the steel flux-return yoke outside the solenoid.
A more detailed description of the CMS detector, together with a definition of the coordinate system,
can be found in Ref.~\cite{Chatrchyan:2008zzk}.
The particle-flow (PF) algorithm~\cite{Sirunyan:2017ulk} aims to reconstruct and identify
each individual particle in an event, using an optimized combination of information from the
various elements of the CMS detector.
The candidate vertex with the largest sum of the square of the transverse momenta $\pt^2$ of the
physics objects is taken to be the primary \pp interaction vertex;
more details are given in Section~9.4.1 of Ref.~\cite{CMS-TDR-15-02}.
From PF candidates, jets are reconstructed using the
\antikt~\cite{Cacciari:2008gp} or the XCone~\cite{Stewart:2015waa} algorithm as implemented in the
\FASTJET software package~\cite{Cacciari:2011ma}. The \antikt jets are obtained using a
distance parameter of 0.4. In the jet clustering procedure, charged PF candidates
are excluded if they are associated to vertices from additional inelastic
\pp interactions within the same bunch crossing (pileup).

The \POWHEG~\cite{Nason:2004rx,Frixione:2007vw,Alioli:2010xd,Frixione:2007nw,
Alioli:2009je,Re:2010bp} v2 generator is used for simulating
\ttbar production at next-to-leading order (NLO).
Alternatively, \ttbar production is simulated with \MGvATNLO v2.2.2~\cite{Alwall:2014hca, Frixione:2002ik} at NLO to
check a potential generator dependence of the measured cross sections.
Background events resulting from the production of single
top quarks are also generated in \POWHEG at NLO,
where spin correlations are taken into account~\cite{Artoisenet:2012st}.
The production of a {\PW} boson with additional jets is simulated using \MGvATNLO at NLO.
Events from Drell--Yan (DY) production with additional jets are simulated
in \MGvATNLO at leading order (LO) and are normalized to the
next-to-next-to-leading-order cross section~\cite{Li:2012wna}.
The simulation of the production of two heavy gauge bosons with additional jets is performed at LO
with \PYTHIA v8.212~\cite{Sjostrand:2014zea}.
Events in which jets are produced only through QCD interactions are also simulated with \PYTHIA at LO.

In simulated \MGvATNLO events the matrix element (ME) calculations at NLO and LO accuracy
are matched to parton showers with the
FxFx~\cite{Frederix:2012ps} and MLM~\cite{Alwall:2007fs} algorithms, respectively.
The parton shower, hadronization process, and multiple-parton interactions (MPI) are simulated using \PYTHIA.
The NNPDF3.0~\cite{Ball:2014uwa} PDFs
at LO and NLO are used for the respective processes simulated at LO and NLO. The UE tune
CUETP8M2T4~\cite{Sirunyan:2019dfx} is used to simulate \ttbar and single top
quark production in the $t$ channel; all other
processes are simulated using
CUETP8M1~\cite{Khachatryan:2015pea,Skands:2014pea}.
The detector response is simulated with the
\GEANTfour package~\cite{Agostinelli:2002hh, Allison:2006ve}. Simulated
events are processed through the software chain used for
collision data and are reweighted to match the observed distribution
in the number of pileup interactions in data.

This analysis uses data recorded with the CMS detector that
correspond to an integrated luminosity of 35.9\fbinv~\cite{CMS-PAS-LUM-17-001}.
Events containing the decay of a top quark to a final state including a muon are selected using a
single-muon trigger~\cite{Khachatryan:2016bia} that requires the
presence of at least one muon candidate with a transverse momentum $\pt>50\GeV$ and $\abs{\eta}<2.4$.
For events containing a final-state electron,
the trigger requires the presence of at least one isolated candidate with
$\pt>27\GeV$, or an electron candidate without an isolation requirement but with $\pt>115\GeV$
and $\abs{\eta}<2.5$, or at least one photon candidate with $\pt>175\GeV$ and
$\abs{\eta}<2.5$.
The latter requirement ensures that events containing electrons with high \pt are selected with
high efficiency.

Lepton candidates (electrons or muons) must have $\pt>55\GeV$, $\abs{\eta}<2.4$.
Following the requirement at the trigger level, electrons with $\pt<120\GeV$ must pass an isolation
requirement~\cite{Sirunyan:2018fpa}, where the isolation is defined as the $\pt$ sum of
charged hadrons and neutral particles in a cone with
radius ${\Delta}R=0.3$ around the electron.
The angular distance between two objects is defined as
${\Delta}R = \sqrt{\smash[b]{(\Delta\eta)^2+(\Delta\phi)^2}}$,
where $\phi$ is the azimuthal angle in radians.
Electrons with $\pt>120\GeV$ and muons with $\pt>55\GeV$ are required to pass a two-dimensional
selection of either ${\Delta}R(\ell,\jet)>0.4$ or $\ptrel(\ell,\jet)>40\GeV$, where \jet is the
\antikt jet with minimal angular separation ${\Delta}R$ from the lepton $\ell$,
and $\ptrel(\ell,\jet)$ is the component of the lepton momentum orthogonal to the \antikt-jet
axis~\cite{Sirunyan:2018ryr, Sirunyan:2018rfo}.
Each selected event must contain a single lepton.

The XCone jets are obtained through a two-step jet clustering~\cite{Thaler:2015xaa}.
First, the exclusive XCone algorithm is applied with a distance parameter of $R_{\text{jet}}=1.2$
and the specification of returning two jets, corresponding to the two boosted top quarks in the event.
Using the constituents of these two large jets as input,
XCone is run again with the distance parameter $R_{\text{sub}}=0.4$ and
the parameter of the number of subjets in each jet $N_{\text{sub}}=3$.
Subjets are only considered if they are within $\abs{\eta}<2.4$.
This procedure results in exactly two large-radius XCone jets with three XCone subjets each.
The final result is not influenced by the number of
subjets within the large XCone jet including the lepton, where $N_{\text{sub}}=2$
would be the natural choice for clustering the visible products of the
decay $\cPqt \to \cPqb \PW \to \cPqb \ell \nu$.
The four-momentum of the lepton candidate is subtracted from the four-momentum of the \antikt jet
or XCone subjet if ${\Delta}R(\ell,\jet)<0.4$.
Jet energy corrections~\cite{Khachatryan:2016kdb} derived for \antikt jets
are applied to \antikt jets and XCone subjets.
The jet energy resolution in simulated events is smeared to match the resolution in data.
An additional correction applied to the XCone-subjet momenta is obtained from simulated \ttbar
events in the all-jets channel to account for differences between the XCone-subjet momenta
and the momenta of \antikt jets.
This correction is parametrized as a function of XCone subjet \pt and $\abs{\eta}$, and has an average
size of 2\%, with an average uncertainty of 0.3\%.

The four-momenta of the three XCone subjets are combined to form the final XCone jet.
The XCone jet used to perform the measurement is the one with the largest
distance ${\Delta}R$ to the selected lepton.
Each of the three XCone subjets in this jet must have $\pt>30\GeV$.
The XCone-jet mass \mjet is the invariant mass of
all PF candidates clustered into the three XCone subjets.

In order to identify jets originating from the hadronization of \cPqb quarks, the combined secondary
vertex v2 (CSVv2)~\cite{Sirunyan:2017ezt} algorithm is applied to the \antikt jets.
These candidate \cPqb jets are required to have $\pt>30\GeV$ and $\abs{\eta}<2.4$,
and must pass the tight working point of the CSVv2 algorithm.

The fiducial region chosen for this measurement is studied using simulations at the
particle level, defined by all particles with average lifetimes longer than $10^{-8}$\unit{s}.
The kinematic phase space of this region is defined through \ttbar events containing one lepton
with $\ptlep>60\GeV$, which originates from the decay of a $\PW$ boson;
the $\tau$ lepton decays are not considered part of the signal.
Particle-level jets are obtained with a clustering identical to the one in data.
The particle-level XCone jet with largest distance ${\Delta}R$ to the lepton
is required to have $\pt>400\GeV$, and each of its XCone subjets must have $\pt>30\GeV$.
Its mass has to be greater than the mass obtained by summing the four-momenta of the
second-highest XCone jet in \pt and the lepton.
The resulting distribution in \mjet at the particle level has a width half as large as for
Cambridge--Aachen (CA) jets~\cite{CACluster1,CACluster2} with
$R_{\text{jet}}=1.2$, as used in a previous measurement~\cite{Sirunyan:2017yar}.
The improvement is due to the two-step XCone jet clustering procedure, which
acts as a grooming algorithm~\cite{Larkoski:2017jix, Marzani:2019hun, Asquith:2018igt},
similar to trimming~\cite{Krohn:2009th}, on the large jet.
The advantage of XCone over other grooming algorithms in this measurement
is its dynamical interpolation between the resolved and boosted regime, \ie,
between three well-separated subjets and three subjets close together, which would not be
resolved by other reconstruction methods.

\begin{figure}[tb]
  \centering
    \includegraphics[width=0.49\textwidth]{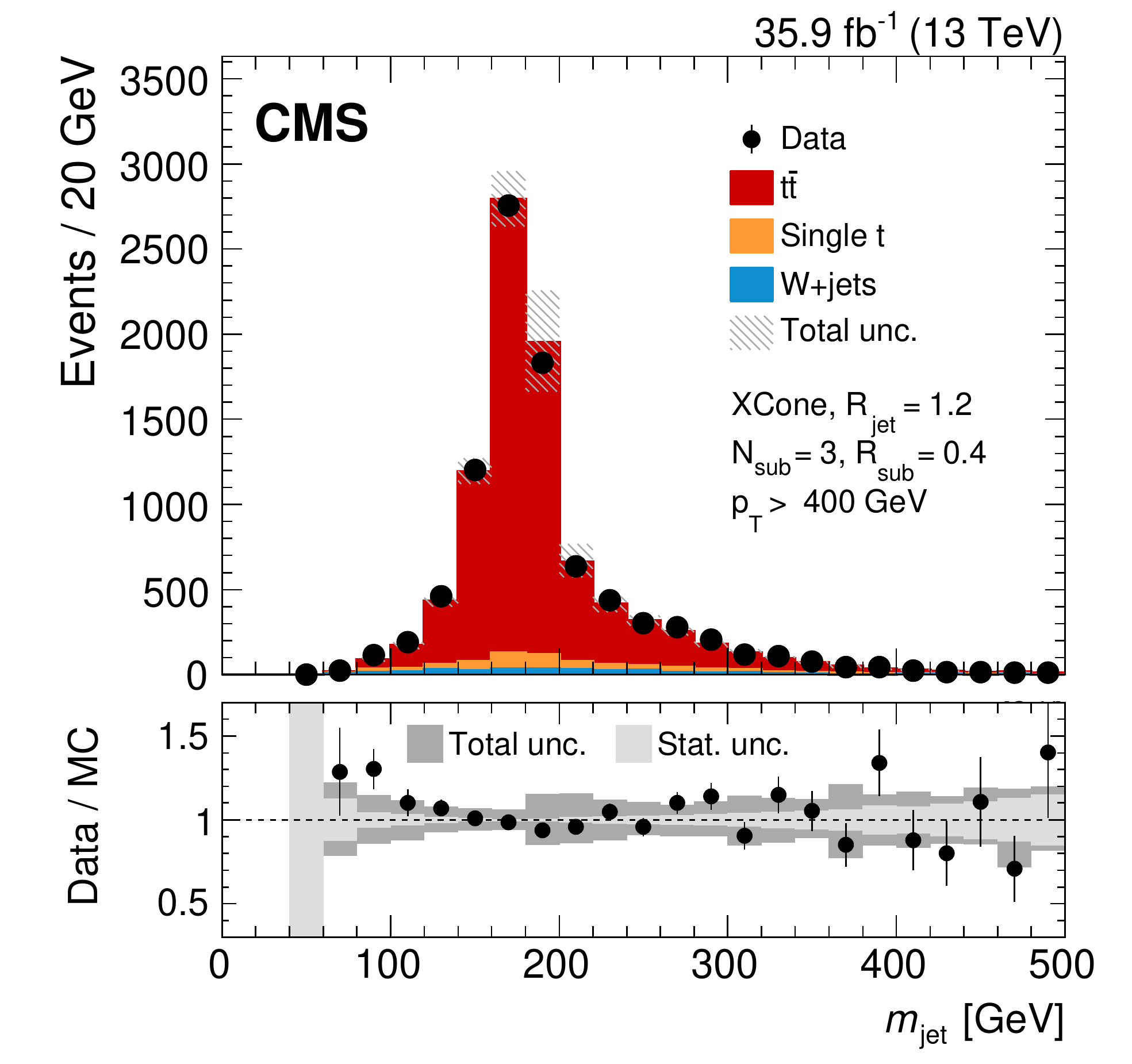}
    \caption{Reconstructed distribution of \mjet after the full event selection in the \ljets channel.
    The vertical bars on the points show the statistical uncertainty.
    The hatched region shows the total uncertainty in the
    simulation, including the statistical and experimental systematic uncertainties. The lower panel
    shows the ratio of the data to the simulation. The uncertainty band includes the statistical and
    experimental systematic uncertainties, where the statistical (light grey) and total (dark grey)
    uncertainties are shown separately in the ratio.
    \label{fig:reco}}
\end{figure}
At the reconstruction level the same criteria are used as in the definition of
the fiducial phase space at the particle level.
In addition, at the reconstruction level an event has to have at least one \cPqb-tagged \antikt jet
and $\ptmiss>50\GeV$, which suppresses non-\ttbar backgrounds.
Here, \ptmiss is the magnitude of the negative vector sum of the
transverse momenta of the PF candidates in an event~\cite{Sirunyan:2019kia}.
The resulting \mjet distribution for XCone jets with $\ptjet>400\GeV$ is displayed in Fig.~\ref{fig:reco}.
Backgrounds originate from singly produced top quarks and from \Wjets events.
Contributions from DY+jets, diboson, and QCD multijet production are found to be negligible.
The \ttbar simulation is scaled, such that the number of
simulated events matches the number of background-subtracted events in data.
The distribution shows a pronounced and narrow peak close to
the value of \mtop.
The XCone-jet reconstruction results in a large improvement of the experimental
resolution in \mjet. With XCone a resolution of 6\% is achieved,
compared to a resolution of approximately 14\% for CA jets with $R_{\text{jet}}=1.2$.

The measurement at the particle level uses a regularized
unfolding procedure based on a least-squares fit, implemented in the
\textsc{TUnfold}~\cite{Schmitt:2012kp} framework.
The optimal regularization strength is determined through a
minimization of the average global correlation coefficient in the output
bins~\cite{Schmitt:2016orm}.
The response matrix is evaluated by using $\ttbar$ events simulated with \POWHEG that
pass the particle- or reconstruction-level requirements.
Prior to the unfolding, contributions from background processes are subtracted from data.
Sideband regions are included in the unfolding process to constrain migrations into and
out of the measurement phase space. Five sideband regions are defined by the requirements:
$55<\ptlep<60\GeV$, $350<\ptjet<400\GeV$, at least one XCone subjet with $\pt<30\GeV$,
\mjet less than the mass of the second XCone jet and lepton system,
and at least one \antikt jet passing a looser \cPqb tagging requirement with no \antikt
jet passing the tight \cPqb tagging requirement.
In addition, the measurement region is divided into three bins in \ptjet.
Except for the sideband with a looser \cPqb tag, all sideband selections
have corresponding selections at the particle level in the evaluation of the migration matrix.
In this matrix, the number of bins in \mjet at the particle level is larger
than the number of bins in which the final measurement is presented.
This helps to reduce the dependence on variations
in signal modeling through a more precise determination of migration effects.
The electron and muon channels are combined before the unfolding
to increase the statistical precision, but are also unfolded separately to verify
their consistency.

Experimental uncertainties are estimated using simulation and propagated through the unfolding process.
We consider uncertainties in the pileup reweighting~\cite{Sirunyan:2018nqx},
trigger, lepton identification and \cPqb tagging~\cite{Sirunyan:2017ezt} efficiencies,
and also those related to the jet energy scale~\cite{Khachatryan:2016kdb} and
jet energy resolution for \antikt jets and XCone subjets, and additional XCone-subjet corrections.
Uncertainties related to the integrated luminosity~\cite{CMS-PAS-LUM-17-001}
and the production cross sections of all significant background
processes~\cite{Khachatryan:2016kzg, Chatrchyan:2014mua, Sirunyan:2016cdg,
Kidonakis:2012rm, Gehrmann:2014fva, Khachatryan:2016tgp}
are also included.
Uncertainties arising from choices in modeling signal include changes made in
renormalization and factorization scales \mur and \muf, changes in \mtop by $\pm3\GeV$,
changes in PDFs, and choices in modeling of parton showers (PS) and their
matching to the ME calculation and the underlying event (UE).
Uncertainties in the modeling of PS include changes in scales of initial- and final-state radiation
(ISR and FSR) and changes in the ME matching parameter $h_\text{damp}$~\cite{Sirunyan:2019dfx}.
The uncertainty related to modeling the UE is estimated by changing the model of color
reconnection in \PYTHIA~\cite{Sjostrand:1987su} and using two other schemes~\cite{Argyropoulos:2014zoa, Christiansen:2015yqa}.
Uncertainties from modeling \cPqb quark fragmentation and the semileptonic
branching fractions of \cPqb hadrons are found to be negligible.

\begin{figure}[tb]
  \centering
    \includegraphics[width=\cmsFigWidth]{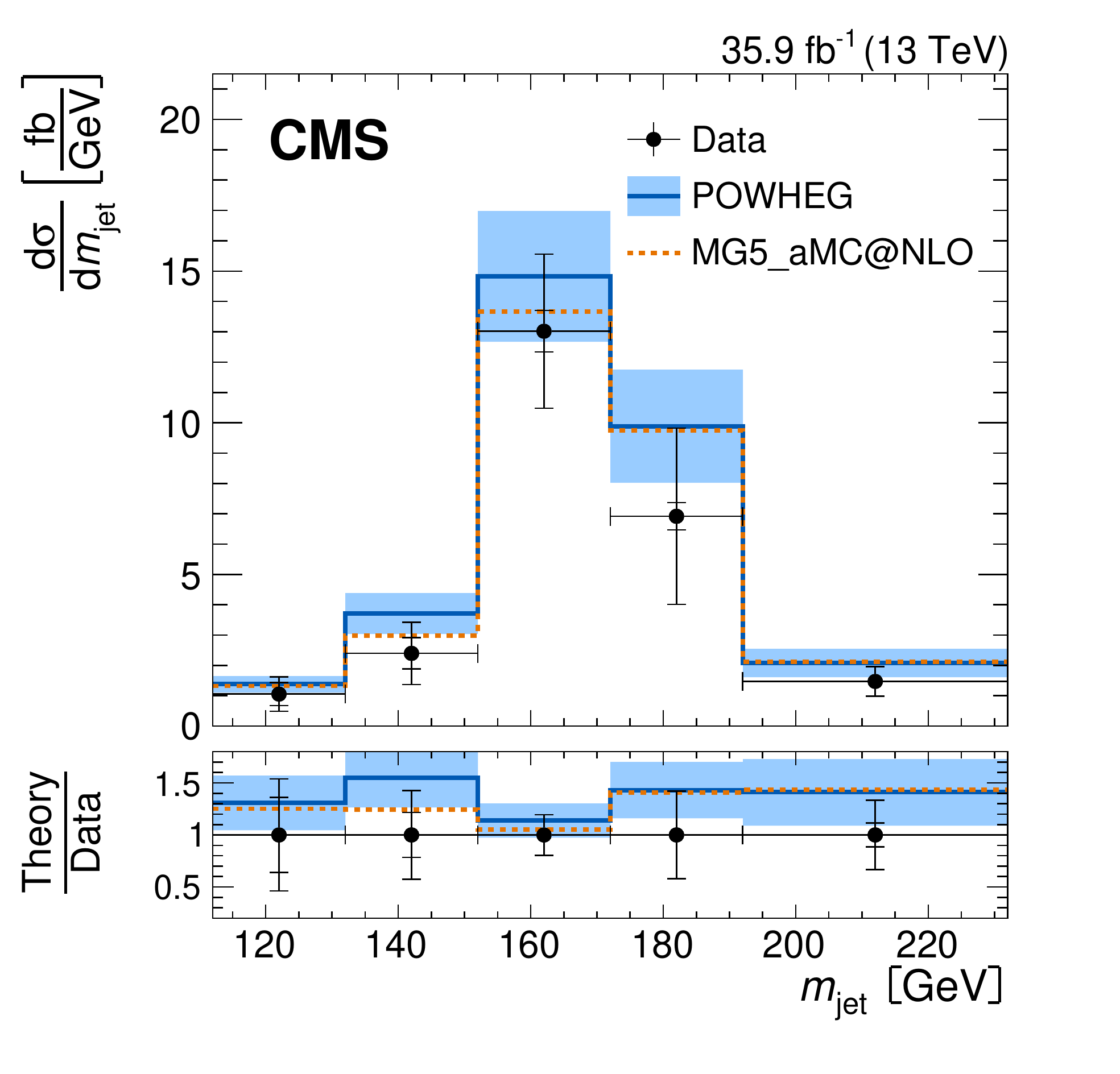}
    \includegraphics[width=\cmsFigWidth]{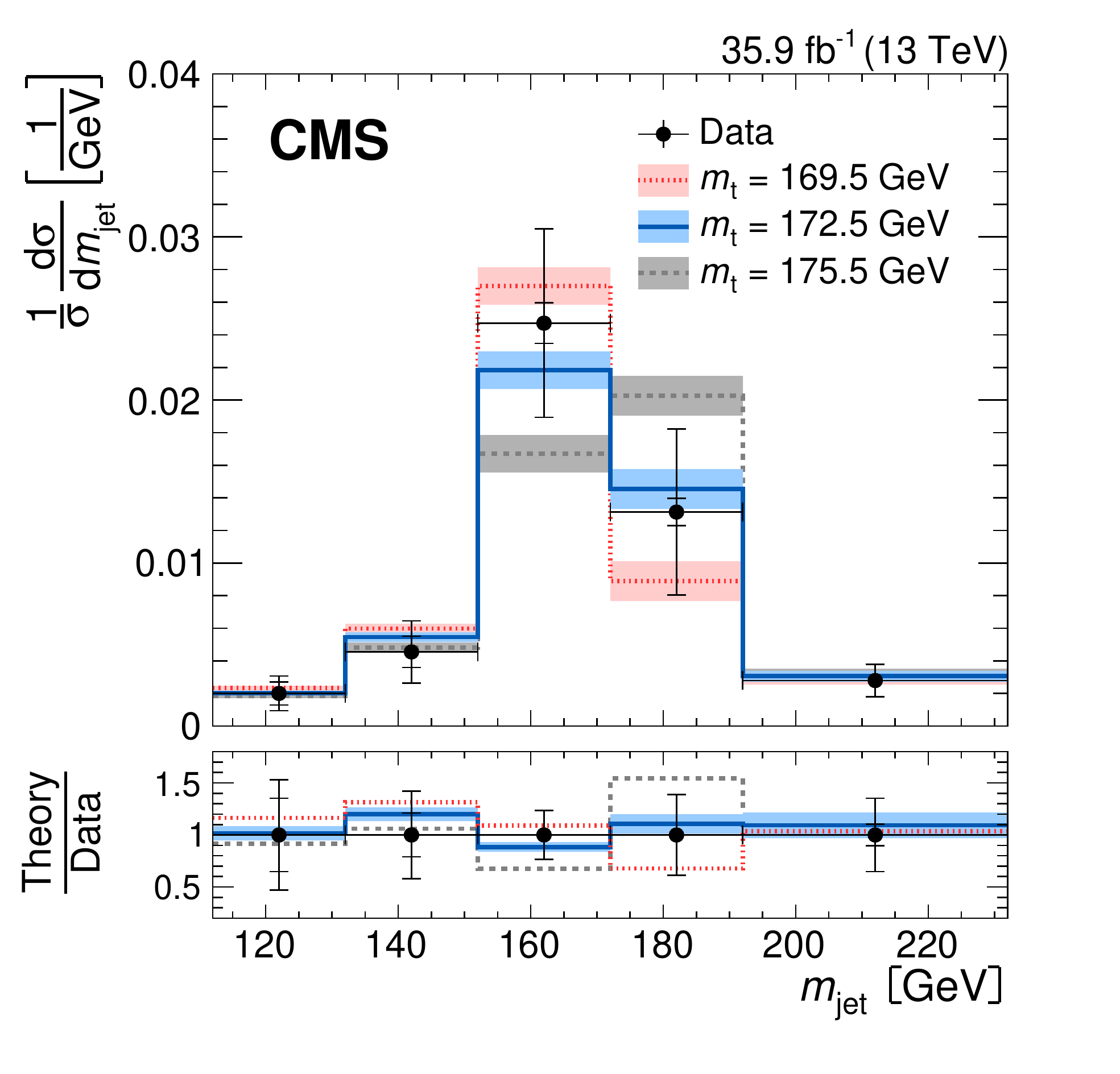}
    \caption{The particle-level \ttbar differential cross section
    in the fiducial region as a function of the XCone-jet mass (\cmsLeft). The measurement is compared to
    predictions from \POWHEG and \MGvATNLO with $\mtop = 172.5\GeV$. Theoretical uncertainties
    are shown as bands for the predictions from \POWHEG.
    The normalized differential cross section (\cmsRight) is compared to predictions from \POWHEG
    for different values of \mtop.
    The vertical bars represent the statistical (inner) and the total (outer) uncertainties.
    The horizontal bars reflect the bin widths.
    The lower panels show the ratios of theoretical predictions to data.
    \label{fig:result}}
\end{figure}

The measured differential cross section in data is shown in Fig.~\ref{fig:result} (\cmsLeft)
and compared to the predictions from \POWHEG and \MGvATNLO with $\mtop = 172.5\GeV$.
In the peak region, the total relative uncertainty is between 16 and 36\%,
of which the dominant contribution is 12--31\% from the jet energy scale uncertainty.
The largest model uncertainty is from FSR modeling, with an uncertainty of 4--18\%.
The statistical uncertainty is 6--7\%.
The total measured \ttbar cross section in the fiducial region of $112<\mjet<232\GeV$ is
$\sigma = 527 \pm 15 \stat \pm 39 \exper \pm 29 \model\fb$.
The cross section predicted by \POWHEG is $680 \pm 109 \fb$, where
the theoretical uncertainty is obtained by changing the scales $\mur$ and $\muf$,
the ISR and FSR PS scales, the parameter $h_\text{damp}$, and the
UE modeling in simulation.
A smaller cross section is observed in data relative to simulation,
in agreement with previous high-\pt top quark
measurements~\cite{Aad:2015hna, Aaboud:2018eqg, Sirunyan:2017yar,
Khachatryan:2016gxp, Sirunyan:2018wem}.

Figure~\ref{fig:result} (\cmsRight) shows the normalized differential cross section as a function of
\mjet, which is obtained by dividing the differential cross section
by the total cross section in the fiducial region.
The normalized differential cross section
benefits from a partial cancellation of systematic uncertainties and shows good agreement
with the prediction from \POWHEG for a value of $\mtop=172.5\GeV$.

The normalized differential cross section can be used to extract a value of \mtop.
A fit is performed based on the $\chi^2$ evaluated as
$\chi^2 = d^T V^{-1} d$,
where $d$ is the vector of differences between the measured normalized cross sections
and the predictions obtained from \POWHEG for different values of \mtop.
The symbol $V$ represents the covariance matrix that contains
statistical, experimental systematic, signal modeling in the unfolding, and theoretical uncertainties.
The result is
\begin{linenomath}
\ifthenelse{\boolean{cms@external}}
{
\begin{equation*}
\begin{aligned}
\mtop &= 172.6 \pm 0.4\stat \pm 1.6\exper\\
      &\pm 1.5\model \pm 1.0\thy\GeV.
\end{aligned}
\end{equation*}
}
{
\begin{equation*}
\begin{aligned}
\mtop &= 172.6 \pm 0.4\stat \pm 1.6\exper \pm 1.5\model \pm 1.0\thy\GeV. \\
\end{aligned}
\end{equation*}
}
\end{linenomath}
This result is a determination of \mtop from decays of boosted top quarks, with an
average energy scale of approximately 480\GeV, much larger than the scale
in \mtop measurements from threshold production.
The improvement in precision by a factor of 3.6 relative to the measurement at 8\TeV~\cite{Sirunyan:2017yar}
is attributed primarily to the novel jet reconstruction using XCone.
The improvement by a factor of two in both, the \mjet width at the particle level
and experimental resolution, together with more integrated luminosity and an increased value of $\sqrt{s}$,
provides a reduction by a factor of about 14 in the statistical uncertainty.

The systematic uncertainties are also reduced through the XCone-jet reconstruction,
which enables a more precise calibration of the XCone-subjet energies and a better stability
against contributions from pileup and the UE.
Uncertainties from modeling are reduced through the use of additional sideband regions
with higher granularity in the unfolding.

{\tolerance=400
In summary, a measurement has been presented of the \ttbar differential cross section for
$\cPqt \to \cPqb \PW \to \cPqb \qqbar'$ decays of boosted top quarks as a function of the jet mass \mjet. 
A determination of \mtop from the normalized \mjet distribution provides a value of
$172.6 \pm 2.5\GeV$, with an uncertainty close to that of events
at the \ttbar production threshold.
This measurement shows for the first time the importance of boosted top quarks
for extracting standard model parameters such as \mtop.
The differential cross section as a function of \mjet will enable a determination of \mtop
using precise analytical calculations,
feasible only in the boosted regime~\cite{Hoang:2017kmk}.
This is an important step in understanding the ambiguities arising
between the top quark pole mass and \mtop measurements at hadron colliders.
The novel reconstruction technique using the XCone jet algorithm results in 
the accuracy necessary for precision measurements at large top quark momenta, 
which will become increasingly important in future work at the LHC.
\par}

\begin{acknowledgments}
We congratulate our colleagues in the CERN accelerator departments for the excellent performance of the LHC and thank the technical and administrative staffs at CERN and at other CMS institutes for their contributions to the success of the CMS effort. In addition, we gratefully acknowledge the computing centers and personnel of the Worldwide LHC Computing Grid for delivering so effectively the computing infrastructure essential to our analyses. Finally, we acknowledge the enduring support for the construction and operation of the LHC and the CMS detector provided by the following funding agencies: BMBWF and FWF (Austria); FNRS and FWO (Belgium); CNPq, CAPES, FAPERJ, FAPERGS, and FAPESP (Brazil); MES (Bulgaria); CERN; CAS, MoST, and NSFC (China); COLCIENCIAS (Colombia); MSES and CSF (Croatia); RPF (Cyprus); SENESCYT (Ecuador); MoER, ERC IUT, PUT and ERDF (Estonia); Academy of Finland, MEC, and HIP (Finland); CEA and CNRS/IN2P3 (France); BMBF, DFG, and HGF (Germany); GSRT (Greece); NKFIA (Hungary); DAE and DST (India); IPM (Iran); SFI (Ireland); INFN (Italy); MSIP and NRF (Republic of Korea); MES (Latvia); LAS (Lithuania); MOE and UM (Malaysia); BUAP, CINVESTAV, CONACYT, LNS, SEP, and UASLP-FAI (Mexico); MOS (Montenegro); MBIE (New Zealand); PAEC (Pakistan); MSHE and NSC (Poland); FCT (Portugal); JINR (Dubna); MON, RosAtom, RAS, RFBR, and NRC KI (Russia); MESTD (Serbia); SEIDI, CPAN, PCTI, and FEDER (Spain); MOSTR (Sri Lanka); Swiss Funding Agencies (Switzerland); MST (Taipei); ThEPCenter, IPST, STAR, and NSTDA (Thailand); TUBITAK and TAEK (Turkey); NASU (Ukraine); STFC (United Kingdom); DOE and NSF (USA). \end{acknowledgments}

\bibliography{auto_generated}
\cleardoublepage \appendix\section{The CMS Collaboration \label{app:collab}}\begin{sloppypar}\hyphenpenalty=5000\widowpenalty=500\clubpenalty=5000\vskip\cmsinstskip
\textbf{Yerevan Physics Institute, Yerevan, Armenia}\\*[0pt]
A.M.~Sirunyan$^{\textrm{\dag}}$, A.~Tumasyan
\vskip\cmsinstskip
\textbf{Institut f\"{u}r Hochenergiephysik, Wien, Austria}\\*[0pt]
W.~Adam, F.~Ambrogi, T.~Bergauer, J.~Brandstetter, M.~Dragicevic, J.~Er\"{o}, A.~Escalante~Del~Valle, M.~Flechl, R.~Fr\"{u}hwirth\cmsAuthorMark{1}, M.~Jeitler\cmsAuthorMark{1}, N.~Krammer, I.~Kr\"{a}tschmer, D.~Liko, T.~Madlener, I.~Mikulec, N.~Rad, J.~Schieck\cmsAuthorMark{1}, R.~Sch\"{o}fbeck, M.~Spanring, D.~Spitzbart, W.~Waltenberger, C.-E.~Wulz\cmsAuthorMark{1}, M.~Zarucki
% ...
\vskip\cmsinstskip
\dag: Deceased\\
1:  Also at Vienna University of Technology, Vienna, Austria\\
% ...
\end{sloppypar}
\end{document}